\documentclass[a4paper]{jpconf}
\usepackage{graphicx}
\begin{document}
\title{Pure Neutron Matter Constraints and Nuclear Symmetry Energy}

\author{F J Fattoyev$^{1,2, \ast}$, W G Newton$^{1, \dag}$, Jun Xu$^{1, \ddag}$ and Bao-An Li$^{1,3, \S}$}

\address{$^{1}$Department of
Physics and Astronomy, Texas A\&M University-Commerce, Commerce,
Texas 75429-3011, USA \\ $^{2}$Institute of Nuclear Physics,
Tashkent 100214, Uzbekistan \\ $^{3}$Department of Applied Physics,
Xian Jiao Tong University, Xian 710049, China}

\ead{$^{\ast}$Farrooh.Fattoyev@tamuc.edu,
$^{\dag}$William.Newton@tamuc.edu, $^{\ddag}$Jun.Xu@tamuc.edu,
$^{\S}$Bao-An.Li@tamuc.edu}

\begin{abstract}
In this review, we will discuss the results of our recent
work~\cite{Fattoyev:2012b} to study the general optimization of the
pure isovector parameters of the popular relativistic mean-field
(RMF) and Skyrme-Hartree-Fock (SHF) nuclear energy-density
functionals (EDFs), using constraints on the pure neutron matter
(PNM) equation of state (EoS) from recent {\sl ab initio}
calculations. By using RMF and SHF parameterizations that give
equivalent predictions for ground-state properties of doubly magic
nuclei and properties of symmetric nuclear matter (SNM) and PNM, we
found that such optimization leads to broadly consistent symmetry
energy $J$ and its slope parameter $L$ at saturation density within
a tight range of $\sigma(J) < 2$ MeV and $\sigma(L) < 6$ MeV. We
demonstrate that a clear model dependence shows up (a) in the
curvature parameter of the symmetry energy $K_{\rm sym}$, (b) the
symmetry energy at supra-saturation densities, and (c) the radius of
neutron stars.
\end{abstract}

\section{Introduction}

Phenomenological nuclear effective interactions offer a compact
description of the \emph{in-medium} nucleon-nucleon interaction and
are useful tools in the applications of both the nuclear structure
and the astrophysical phenomena. The effective interaction is
typically dependent on few parameters representing, for example,
coupling constants, which are often fit to well-determined
experimental nuclear observables such as binding energies, charge
radii, single particle energy spectra and spectra of collective
excitations. One of the main objective of modern nuclear many-body
theory is to obtain an EDF~\cite{UNEDF} with clear physical
connections to \emph{ab initio} nucleon-nucleon interactions and
QCD.

In the recent years, much effort has been devoted to constrain the
energy per neutron of PNM ($E_{\rm PNM}$) at sub-saturation
densities. By studying the universal behavior of resonant Fermi
gases with infinite scattering length, a significant constraint is
achieved for the EoS of dilute neutron matter~\cite{Schwenk:2005ka}.
These calculations have been extended to higher densities using the
full power of quantum Monte Carlo methods~\cite{Gezerlis:2009iw,
Gezerlis:2011ai}. Moreover, by studying the physics of chiral
three-nucleon forces the EoS of PNM is obtained perturbatively up to
nuclear saturation density~\cite{Hebeler:2009iv}. Finally, the
auxiliary field diffusion Monte Carlo (AFDMC) technique, which takes
into account the realistic nuclear Hamiltonian containing modern
two- and three-body interactions of the Argonne potential and Urbana
family of three-body nucleon forces, is used to calculate the EoS of
PNM up to and above saturation density~\cite{Gandolfi:2009fj,
Gandolfi:2009nq, Gandolfi:2011xu}.

In this work we concentrate on the widely used
RMF~\cite{Serot:1984ey, Serot:1997xg} and SHF~\cite{Skyrme:1956zz,
Vautherin:1971aw} models, with the latter thought of as a
non-relativistic expansion of the former~\cite{Reinhard:1989zi,
Sulaksono:2003re}. Both models have less than ten free parameters in
their simplest forms. Although there are more than $200$ parameter
sets in the literature for the SHF model~\cite{Dutra:2012mb} and
dozens of parameterizations exist in the simplest form of the RMF
model, e.g.~\cite{Chen:2007ih}, many of these sets are old and have
been superseded by parameter sets fit to more recent and accurate
data. Since the number of experimental observables is usually larger
than the number of free parameters, the problem of optimizing these
EDFs is generally overdetermined, and this results in a significant
degeneracy among parameter sets. Fortunately, one can use the
covariance analysis techniques~\cite{Reinhard:2010wz,
Fattoyev:2011ns} to study correlations between predicted observables
from a particular EDF in its model space. We use the linear
regression method to optimize the two pure isovector parameters of
RMF and SHF models by using the results from the \emph{ab initio}
theoretical calculations of the PNM EoS as our `experimental'
constraints. To obtain meaningful theoretical uncertainties for the
model parameters, as well as for the predicted observables, we
employ the covariance analysis technique. Optimizing the two pure
isovector parameters will ensure that the predictions for the
well-determined isoscalar observables such as binding energies
$B(A)$ and charge radii $R_{\rm ch}$ of doubly magic nuclei will not
be affected. We find the best fit values and 1$\sigma$ confidence
intervals on the properties of isospin-asymmetric nuclear matter,
such as the symmetry energy parameters. In addition, poorly
constrained observables such as the neutron skin thickness of lead
and neutron-star radii are predicted from the resulting constraints.
We discuss the manifestation of the model dependence in our results,
by exploring the symmetry energy at supra-saturation densities and
properties of neutron stars. We should note that our aim is not to
establish new parameterizations of these EDFs, or to set absolute
constraints on symmetry energy, but to explore as far as possible
the generic constraints that can be placed by each model on
neutron-rich systems once constrained by the information from the
PNM EoS.

\section{Nuclear Symmetry Energy}

The nuclear symmetry energy, $S(\rho)$, is defined as the
coefficient of the leading term of isospin asymmetry parameter,
$\alpha=(\rho_{\rm n}-\rho_{\rm p})/\rho$, in the expression of the
binding energy per nucleon in neutron-rich nuclear matter
\begin{equation}\label{eanm}
E(\rho,\alpha) = E_0(\rho) + S(\rho) \alpha^2 +
\mathcal{O}(\alpha^4) \ ,
\end{equation}
where $\rho$ is the baryon number density with $\rho_{\rm n}$
($\rho_{\rm p}$) being the neutron (proton) number density. Around
the saturation density $\rho_0$, one can express the symmetry energy
as
\begin{equation}
S(\rho) = J + L \chi + \frac{1}{2}K_{\rm sym}\chi^2 +
\mathcal{O}(\chi^3) \ ,
\end{equation}
where $\chi \equiv \left(\rho - \rho_0\right)/3\rho_0$, $J$ is the
value of the symmetry energy at saturation density, $L$ is the slope
parameter, and $K_{\rm sym}$ is the curvature parameter of the
symmetry energy at
saturation density. % given, respectively, by the following
%expressions:
%\begin{eqnarray}
%&&  L = 3 \rho_0 \left(\frac{\partial S(\rho)}{\partial
%\rho}\right)_{\rho=\rho_0} \ , \\
%&& K_{\rm sym} = 9 \rho_0^2 \left(\frac{\partial^2 S(\rho)}{\partial
%\rho^2}\right)_{\rho=\rho_0} \ .
%\end{eqnarray}
The coefficients of the higher-order terms in Eq.~(\ref{eanm}) are
generally much smaller than $S(\rho)$, so it is usually a good
approximation to write the symmetry energy as the difference between
the energy per nucleon of PNM and SNM, i.e., $S(\rho) \approx
E(\rho, 1) - E_0(\rho)$. However, in this work we will not use such
an approximation, but rather calculate it from the full analytical
expression in a given model.

\section{Relativistic Mean-Field and Skyrme-Hartree-Fock Models}

We apply the constraints on the microscopic PNM calculations to the
two popular phenomenological nuclear many-body models to study the
nuclear symmetry energy and related quantities of nuclear physics
and nuclear astrophysics. The commonly used RMF model contains an
isodoublet nucleon field ($\psi$) interacting via the exchange of
the scalar-isoscalar $\sigma$-meson ($\phi$), the vector-isoscalar
$\omega$-meson ($V^{\mu}$), the vector-isovector $\rho$-meson (${\bf
b}^{\mu}$), and the photon ($A^{\mu}$)~\cite{Serot:1984ey,
Serot:1997xg, Mueller:1996pm}. The effective Lagrangian density for
the model can be written as
\begin{eqnarray}
{\mathcal L} &=& \bar\psi \left[\gamma^{\mu} \left(i \partial_{\mu}
\!-\! g_{\rm v}V_\mu  \!-\! \frac{g_{\rho}}{2}{\mbox{\boldmath
$\tau$}}\cdot{\bf b}_{\mu} \!-\!
\frac{e}{2}(1\!+\!\tau_{3})A_{\mu}\right)  \!-\! \left(M \!-\!
g_{\rm s}\phi\right) \right]\psi +
\frac{1}{2}\partial_{\mu}\phi\,\partial^{\mu} \phi
-\frac{1}{2}m_{\rm s}^{2}\phi^{2} \nonumber \\
&-& \frac{1}{4}V^{\mu\nu}V_{\mu\nu} + \frac{1}{2}m_{\rm
v}^{2}V^{\mu}V_{\mu} - \frac{1}{4}{\bf b}^{\mu\nu}\cdot{\bf
b}_{\mu\nu} + \frac{1}{2}m_{\rho}^{2}\,{\bf b}^{\mu}\cdot{\bf
b}_{\mu} -\frac{1}{4}F^{\mu\nu}F_{\mu\nu} -
          U(\phi,V_{\mu},{\bf b_{\mu}}) \;,
 \label{LDensity}
\end{eqnarray}
where $V_{\mu\nu}\equiv \partial_{\mu}V_{\nu} -
\partial_{\nu}V_{\mu}$, ${\bf b}_{\mu\nu} \equiv \partial_{\mu}{\bf b}_{\nu}
 - \partial_{\nu}{\bf b}_{\mu}$, and $F_{\mu\nu} \equiv \partial_{\mu}A_{\nu} - \partial_{\nu}A_{\mu}$ are the
isoscalar, isovector, and electromagnetic field tensors,
respectively. The nucleon mass $M$ and meson masses $m_{\rm s}$,
$m_{\rm v}$, and $m_{\rho}$ may be treated (if wished) as empirical
parameters. The effective potential $U(\phi,V_{\mu},{\bf b_{\mu}})$
consists of non-linear meson interactions that simulates the
complicated dynamics encoded in few model parameters. In the present
work we use the following form of the effective
potential~\cite{Todd-Rutel:2005fa}:
%%%
\begin{equation}
  U(\phi,V^{\mu},{\bf b}^{\mu})  =
    \frac{\kappa}{3!} (g_{\rm s}\phi)^3 \!+\!
    \frac{\lambda}{4!}(g_{\rm s}\phi)^4 \!-\!
    \frac{\zeta}{4!}   g_{\rm v}^4(V_{\mu}V^\mu)^2 -
   \Lambda_{\rm v} g_{\rho}^{2}\,{\bf b}_{\mu}\cdot{\bf b}^{\mu}
           g_{\rm v}^{2}V_{\nu}V^\nu\;.
\label{USelf}
\end{equation}
%%%
This model is described by seven interaction parameters: $\{g_{\rm
s}, g_{\rm v}, g_{\rho}, \kappa, \lambda, \zeta, \Lambda_{\rm v}\}$.
%Note that power counting suggests that a consistent Lagrangian
%density should include all terms up to fourth order in the meson
%fields. However, the existing database of both laboratory and
%observational data appears to be accurately described by the the
%minimal set of parameters~\cite{Lalazissis:1996rd, Lalazissis:1999,
%Todd-Rutel:2005fa}. Indeed, it was shown that ignoring a subset of
%model parameters that are of the same order in a power-counting
%scheme does not compromise the quality of the
%fit~\cite{Mueller:1996pm, Furnstahl:1996wv}.

The standard form of the energy density obtained from the zero-range
Skyrme interaction using the Hartree-Fock method can be written
as~\cite{Chabanat:1997qh}
\begin{eqnarray}
{\mathcal H} &=& \frac{\hbar^2}{2M} \tau  \!+\!
t_0\left[\left(2+x_0\right)\rho^2 -
\left(2x_0+1\right)\left(\rho_{\rm n}^2 + \rho_{\rm
p}^2\right)\right]/4 \nonumber \\ &+&
t_3\rho^{\sigma}\left[\left(2+x_3\right)\rho^2 -
\left(2x_3+1\right)\left(\rho_{\rm n}^2 + \rho_{\rm
p}^2\right)\right]/24 \nonumber \\ &+& \left[t_2\left(2x_2+1\right)
-t_1\left(2x_1+1\right)\right]\left(\tau_n\rho_n +
\tau_p\rho_p\right)/8 + \left[t_1\left(2+x_1\right)
+t_2\left(2+x_2\right)\right]\tau\rho/8 \nonumber \\ &+&
\left[3t_1\left(2+x_1\right)
-t_2\left(2+x_2\right)\right]\left(\nabla \rho\right)^2/32 -
\left[3t_1\left(2x_1+1\right)+t_2\left(2x_2+1\right)\right]\left[\left(\nabla
\rho_{\rm n}\right)^2 + \left(\nabla \rho_{\rm p}\right)^2\right]/32
\nonumber \\ &+& W_0\left[\vec{J} \cdot \nabla\rho  + \vec{J}_{\rm
n} \cdot \nabla\rho_{\rm n} + \vec{J}_{\rm p} \cdot \nabla\rho_{\rm
p}\right]/2 + \left(t_1-t_2\right)\left[J_{\rm n}^2 + J_{\rm
p}^2\right]/16 -\left(t_1x_1 + t_2x_2\right)J^2/16 \ .
\end{eqnarray}
Here $\rho_q$, $\tau_q$, and $\vec{J}_q$ ($q={\rm p}, {\rm n}$) are,
respectively, the number, kinetic energy, and spin-current
densities, and $\rho$, $\tau,$ and $\vec{J}$ are the corresponding
total densities. The SHF model is expressed in terms of nine Skyrme
parameters: $\{t_0, t_1, t_2, t_3, x_0, x_1, x_2, x_3, \sigma\}$ and
the spin-orbit coupling constant $W_0$, which is taken as $133.3$
${\rm MeV} \, {\rm fm}^5$~\cite{Chen:2010qx} in the present work.

\section{Linear Regression and Covariance Analysis Method}

The linear regression and covariance analysis method is discussed in
details in Ref.~\cite{Brandt:1999} and its power has been recently
illustrated in Refs.~\cite{Reinhard:2010wz, Fattoyev:2011ns,
Fattoyev:2012rm}. In a nutshell, one can describe it as follows.
First, an optimal parameters set is found for a given
phenomenological model $\mathcal{O}_{n}^{\rm (th)}({\bf p})$ as a
function of the $F$ model parameters ${\bf
p}\!=\!(p_{1},\ldots,p_{F})$, by minimizing the quality measure
$\chi^2$ through a method of a least-squares fit:
%%%
\begin{equation}
 \chi^{2}({\bf p}) \equiv \sum_{n=1}^{N}
 \left(\frac{\mathcal{O}_{n}^{\rm (th)}({\bf p})-
 \mathcal{O}_{n}^{\rm (exp)}}
 {\Delta\mathcal{O}_{n}}\right)^{2} \;,
 \label{ChiSquare}
\end{equation}
where $N$ is the number of {\sl experimental} observables
$\mathcal{O}_{n}^{\rm (exp)}$ that are determined with an accuracy
of $\Delta\mathcal{O}_{n}$. Then one can compute the symmetric
matrix of second derivatives:
%%%
\begin{equation}
 \chi^2({\bf p}) -\chi^{2}({\bf p}_{0})
 \equiv \Delta\chi^2({\bf x}) =
 {\bf x}^{T}{\hat{\mathcal M}}\,{\bf x} \;,
 \label{Taylor2}
\end{equation}
%%%
where $ x_{i} \equiv \frac{({\bf p}-{\bf p}_{0})_{i}}{({\bf
p}_{0})_{i}}$. The matrix ${\hat{\mathcal M}}$ contains all the
information about the behavior of the $\chi^2$ function around the
minimum. In particular, one can find the meaningful theoretical
uncertainties by computing the statistical covariance of two
observables $A$ and $B$ defined as follows:
%%%
\begin{equation}
 {\rm cov}(A,B) = \sum_{i,j=1}^{F}
 \frac{\partial A}{\partial x_{i}}
  (\hat{{\mathcal M}}^{-1})_{ij}
 \frac{\partial B}{\partial x_{j}}  \;.
 \label{Covariance}
\end{equation}
The variance $\sigma^{2}(A)$ of a given observable $A$ is then
simply given by $\sigma^2(A)\!=\!{\rm cov}(A,A)$. Finally, one can
plot the covariance ellipses between two observables $A$ and $B$ by
diagonalizing the $2 \times 2$ covariance matrix:
\begin{equation}
\hat{{\mathcal C}} = \left(\begin{array}{cc}
{\rm cov}(A,A) & {\rm cov}(A,B) \\
{\rm cov}(B,A) & {\rm cov}(B,B) \end{array}\right)
\end{equation}
The eigenvalues of this covariance matrix represent the semi-major
and semi-minor axes of the covariance ellipse, while the
eigenvectors provide the orientation of the ellipse.

For our set of `experimental' observables $\mathcal{O}_{n}^{\rm
(exp)}$ in the $\chi^2$ input we choose the theoretical microscopic
calculations of the energy per neutron $E_{\rm PNM}$ in the density
range of $0.04 \leq \rho \leq 0.16$ fm$^{-3}$~\cite{Hebeler:2009iv,
Gandolfi:2009fj, Akmal:1998cf}. We restrict our experimental input
to this band only, where the upper bound is the PNM results at
saturation density, as the extension of the PNM calculations to
higher densities using piecewise polytropes in the chiral effective
theory~\cite{Hebeler:2010jx} was shown to allow a huge uncertainty
window in the EoS. Moreover, the symmetry energy coefficients should
only be sensitive to the EoS around the saturation density.

\section{Results}
\label{Results}

%%%%%%%%%%%%%%%%%%%%%%%%%%%%%%%%%%%%%%%%%%%%%%%%%%%%%%%%%%%%%%%%%
\begin{figure}[h]
\vspace{-0.05in}
\begin{center}
\includegraphics[width=0.85\columnwidth,angle=0]{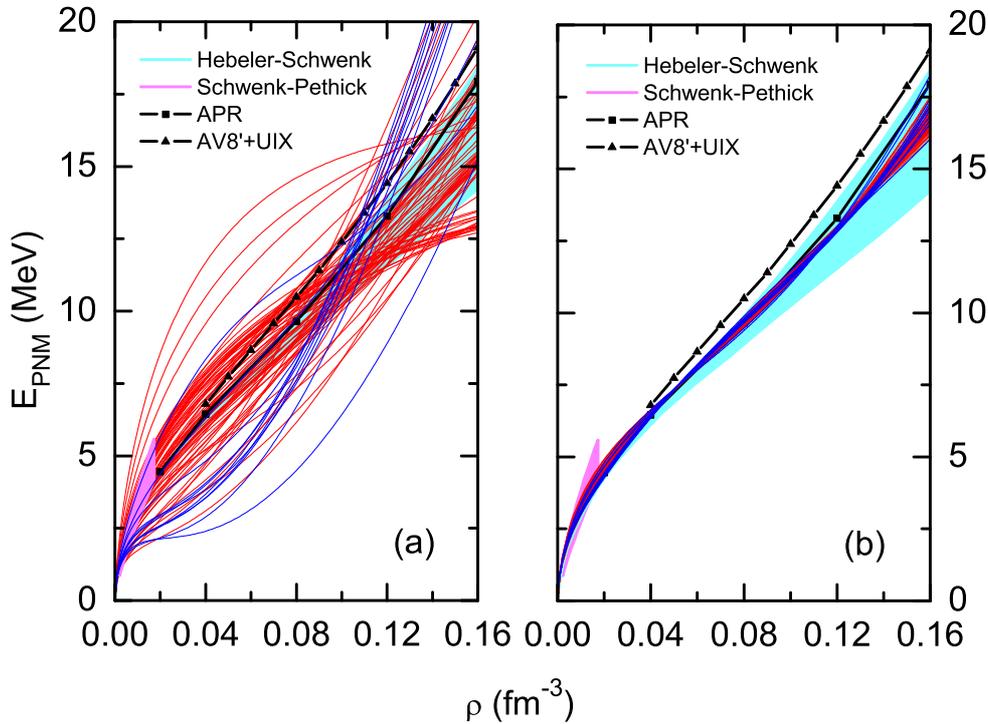}
\caption{Comparing the PNM EoS from 11 RMF (blue lines) and 73 SHF
parameterizations (red lines) that are produced since 1995, with the
AFDMC EoS in the AV8$^{\prime}$+UIX
Hamiltonian~\cite{Gandolfi:2011xu}, the variational APR EoS
~\cite{Akmal:1998cf}, the low-density band from the constraints of
resonant Fermi gases~\cite{Schwenk:2005ka}, and the high-density
band from the chiral effective field theory calculations with
3-neutron forces~\cite{Hebeler:2009iv}, before (a) and after (b) the
PNM optimization.} \label{Fig1}
\end{center}
\end{figure}
%%%%%%%%%%%%%%%%%%%%%%%%%%%%%%%%%%%%%%%%%%%%%%%%%%%%%%%%%%%%%%%%%

We first identify that the two parameters in each model---$g_\rho$
and $\Lambda_{\rm v}$ in the RMF model~\cite{Horowitz:2000xj}, and
$x_0$ and $x_3$ in the SHF model---are {\sl solely} isovector
parameters. The change of these parameters affects only the
isovector sensitive properties of nuclear matter, such as the
symmetry energy $S(\rho)$, while the EoS of SNM and therefore
properties of SNM, such as saturation density $\rho_0$, binding
energy per nucleon at saturation density $E_0$, incompressibility
coefficient at saturation density $K_0$, and isoscalar effective
mass $M^{\ast}$ at saturation all remain unchanged. Therefore we
optimize these two isovector parameters [$F=2$ in
Eqs.~(\ref{Taylor2}) and (\ref{Covariance})] with respect to the
available range of PNM equations of state to constrain the values of
the symmetry energy parameters at saturation density by using the
linear regression and covariance analysis method discussed in the
previous section. Note that most properties of the SNM are
constrained experimentally within less than $10\%$. However, the EoS
of PNM predicted by various parameterizations of both models differ
significantly. Although some of them fall within the band of the
microscopic PNM calculations, for most of the parameterizations
there is almost no or very little agreement with these calculations
as can be seen from Fig. 1 (a). Therefore, in general, the RMF and
the SHF model predictions for the symmetry energy parameters are
substantially different. Once the two isovector parameters are
optimized to the energy per neutron $E_{\rm PNM}$ predictions at
sub-saturation densities [See Fig. 1 (b)], we find that the symmetry
energy parameters at saturation can also be significantly
constrained (See Table \ref{Table1}).

%%%%%%%%%%%%%%%%%%%%%%%%%%%%%%%%%%%%%%%%%%%%%%%%%%%%%%%%%%%%%%%%%
\begin{table}[t]
\caption{Predicted ranges for symmetry energy parameters within RMF
and SHF models before (with superscript `0') and after (without
superscript `0') their pure isovector parameters optimized to PNM,
and taking into account all remaining variation from
parameterizations constructed since 1995.} \label{Table1}
\vspace{+0.05in}
\begin{center}
\begin{tabular}{|l||c|c|c|c|c|c|}
 \hline
 & $J^0$~(MeV) & $J$~(MeV) & $L^0$~(MeV) & $L$~(MeV) & $K_{\tau}^0$~(MeV) & $K_{\tau}$~(MeV) \\
\hline \hline
RMF  & 30.3 -- 38.7 & 30.2 -- 31.4 & 47.2 -- 122.7 & 36.1 -- 59.3 & -701.7 -- -195.3& -329.7 -- -215.7  \\
SHF  & 27.8 -- 39.6 & 30.1 -- 33.2 & \phantom{+}5.8 -- 100.1 & 28.5 -- 64.4 & -514.8 -- -266.3 & -418.8 -- -235.3   \\
\hline
\end{tabular}
\end{center}
\end{table}
%%%%%%%%%%%%%%%%%%%%%%%%%%%%%%%%%%%%%%%%%%%%%%%%%%%%%%%%%%%%%%%%%

To further assess theoretical uncertainties in the symmetry energy
parameters, we select the accurately-calibrated
NL3$^{\ast}$~\cite{Lalazissis:2009zz} and the recent
IU-FSU~\cite{Fattoyev:2010mx} parametrizations from the RMF model.
The IU-FSU is the recent parametrization that predicts a very soft
symmetry energy, and was validated against experimental,
observational, and theoretical data, while the accurately-calibrated
NL3$^{\ast}$ parametrization gives a much stiffer EoS of both SNM
(larger value of $K_0$ and smaller value of $\zeta$ parameter) and a
stiff symmetry energy (larger values of symmetry energy $J$ and
slope $L$) and therefore offers a suitable contrast to IU-FSU.

To compare the RMF and SHF models on the same basis, we create two
SHF parametrizations which give the same properties of nuclear
matter at saturation as the two RMF parametrizations (See Table
\ref{Table2}), through the method of writing the Skyrme parameters
as functions of macroscopic nuclear quantities~\cite{Chen:2010qx,
Chen:2009wv}. These new parametrizations are referred to as
SkNL3$^{\ast}$ and SkIU-FSU forces~\cite{Fattoyev:2012b}.

%%%%%%%%%%%%%%%%%%%%%%%%%%%%%%%%%%%%%%%%%%%%%%%%%%%%%%%%%%%%%%%%%
\begin{table}[h]
\caption{Macroscopic quantities from four reference
parameterizations. They are the nuclear saturation density $\rho_0$,
the binding energy per nucleon $E_0$ and incompressibility $K_0$ of
SNM at saturation, the symmetry energy $J$ and its slope parameter
$L$ at saturation, and the nucleon effective mass $M^{\ast}$ at
saturation. For consistency, we present the Lorentz effective mass
of the RMF model, which is set equal to the isovector and isoscalar
effective masses in the SHF model.} \label{Table2} \vspace{+0.05in}
\begin{center}
\begin{tabular}{|l||c|c|c|c|c|c|}
 \hline
 & $\rho_{0}~({\rm fm}^{-3}) $ & $E_{0}$~(MeV)
           & $K_{0}$~(MeV) & $J$~(MeV) & $L$~(MeV) & $M^{\ast}$~($M$)  \\
\hline \hline
NL3$^{\ast}$     & 0.1500  & $-$16.32 & 258.49 & 38.7 & 122.7 & 0.671    \\
SkNL3$^{\ast}$   & 0.1527  & $-$15.76 & 258.49 & 38.7 & 122.7 & 0.671  \\
IU-FSU           & 0.1546  & $-$16.40 & 231.33 & 31.3 & 47.2  & 0.687 \\
SkIU-FSU         & 0.1575  & $-$15.70 & 231.33 & 31.3 & 47.2  & 0.687  \\
\hline
\end{tabular}
\end{center}
\end{table}
%%%%%%%%%%%%%%%%%%%%%%%%%%%%%%%%%%%%%%%%%%%%%%%%%%%%%%%%%%%%%%%%%

Several definitions of the nucleon effective mass exist in the
literature~\cite{vanDalen:2005ns}. In the RMF model the Dirac
effective mass is defined through the scalar part of the nucleon
self-energy in the Dirac equation. It has been well documented that
there is a strong correlation between the Dirac effective nucleon
mass at saturation density $M^{\ast}_{\rm D}$ and the strength of
the spin-orbit force in nuclei~\cite{Serot:1997xg, Reinhard:1989zi,
Gambhir:1989mp, Bodmer:1991hz}. Indeed, one of the most compelling
features of RMF models is the reproduction of the spin-orbit
splittings in finite nuclei. It is shown that models with effective
masses outside the range $0.58 < M^{\ast}_{\rm D}/M < 0.64$ will not
be able to reproduce empirical spin-orbit
couplings~\cite{Furnstahl:1997tk}, when no tensor couplings are
taken into account. On the other hand, the non-relativistic
effective mass parameterizes the momentum dependence of the single
particle potential, which is the result of a quadratic
parametrization of the single particle spectrum. A recent
study~\cite{Dutra:2012mb} puts a bound of $0.69 < M^{\ast}/M < 1.0$
for the non-relativistic effective masses. It has been
argued~\cite{Jaminon:1989wj} that the so-called Lorentz mass
$M^{\ast}_{\rm L}$  should be compared with the non-relativistic
effective mass extracted from analyses carried out in the framework
of nonrelativistic optical and shell models. For consistency, we
choose the effective mass in the SHF parameterizations to be equal
to the Lorenz mass in the RMF parameterizations (See the last column
of Table \ref{Table2}). Since the RMF model we use in this work
gives the same isoscalar and isovector effective masses, we set them
equal in the reference SHF model too. Finally, the isoscalar
parameters of the two reference Skyrme forces are then re-adjusted
to fit the binding energy and charge radius of $^{208}$Pb by
adjusting only the saturation density $\rho_0$ and the binding
energy $E_0$ of SNM. These ensure that the predictions for the
charge radii and binding energies of other doubly closed-shell
nuclides will be within 1-2\% accuracy. In terms of the predicted
values of isoscalar and isovector bulk observables, both
corresponding RMF and SHF models are therefore almost equivalent.

\begin{figure}[h]
\vspace{-0.05in}
\begin{center}
\includegraphics[width=0.85\columnwidth,angle=0]{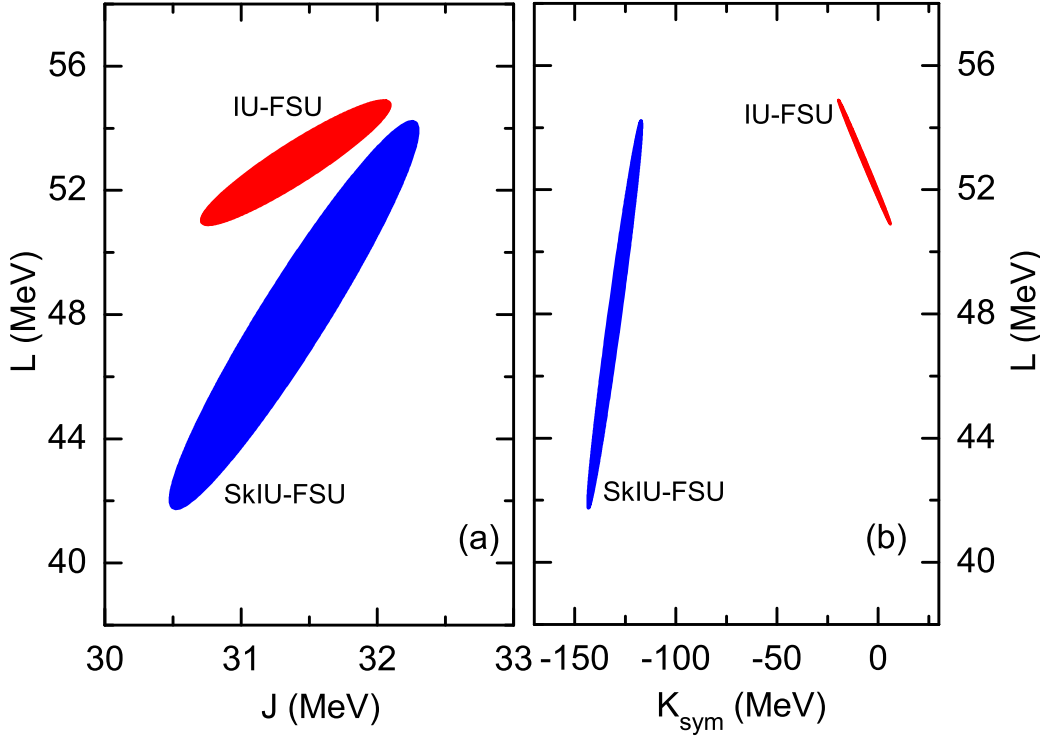}
\caption{$1\sigma$ joint confidence regions for the symmetry energy
$J$ and its slope parameter $L$ (a), and  for the slope parameter
$L$ and curvature parameter $K_{\rm sym}$ (b) of the symmetry energy
at saturation density from the IU-FSU and SkIU-FSU
parameterizations.} \label{Fig2}
\end{center}
\end{figure}

Having obtained the PNM optimized parameter sets, the 1$\sigma$
errors on these two purely isovector parameters can be translated
into equivalent errors on the symmetry energy parameters and the
neutron skin thickness of $^{208}$Pb using the covariance analysis
(See Table \ref{Table3}). The errors in $J$ are less than $\pm 1$
MeV for all the parameterizations. The RMF model gives a relatively
small error in $L$ of around $\pm 2$ MeV, while the SHF model gives
a much larger error around $\pm 6$ MeV. Table \ref{Table3} appears
to indicate that within the $1\sigma$ errors, both models are
consistent in their predicted values of $J$ and $L$. However, a
$1\sigma$ joint confidence regions in the $J$-$L$ plane plotted in
Fig. \ref{Fig2}(a) for both RMF and SHF models shows that in fact
the two models predict non-overlapping regions in $J$-$L$ space.
Both models show a positive correlation between $J$ and $L$, but
with differing slopes. The origin of this difference lies mainly in
the values of the higher-order symmetry energy parameters that are
predicted upon optimization. There is a strong model dependency in
the prediction for the curvature parameter of the symmetry energy
$K_{\rm sym}$ (see Table \ref{Table3}). For example, after the PNM
optimization IU-FSU predicts $K_{\rm sym} = -6.8 \pm 12.9$ MeV,
while its Skyrme-like version predicts a smaller value of $K_{\rm
sym} = -130.2 \pm 13.3$ MeV. When we plot the $1\sigma$ joint
confidence regions in the $K_{\rm sym}$-$L$ plane for both RMF and
SHF models [see  Fig. \ref{Fig2} (b)] further differences can be
seen: there is, generically, a negative correlation between the
slope of the symmetry energy and $K_{\rm sym}$ in the RMF model,
while this correlation is positive in the case of the SHF model.

%%%%%%%%%%%%%%%%%%%%%%%%%%%%%%%%%%%%%%%%%%%%%%%%%%%%%%%%%%%%%%%%%
\begin{table}
\caption{Isovector observables and associated 1$\sigma$ error bars
from four reference parameterizations after the PNM constraints are
applied. Values are shown for the symmetry energy at $\rho = 0.1$
${\rm fm}^{-3}$ $S_{0.1}$ and at saturation density $J$, slope
parameter $L$, curvature parameter $K_{\rm sym}$, isospin-dependent
part of incompressibility $K_{\tau}$, and the neutron skin thickness
$R_{\rm skin}$ of $^{208}$Pb. All the quantities are in MeV apart
from $R_{\rm skin}$, which is in fm.} \label{Table3}
\begin{center}
\begin{tabular}{|l||c|c|c|c|c|c|}
 \hline & $S_{0.1}$ & $J$            & $L$             & $K_{\rm sym}$   &  $K_{\tau}$  &  $R_{\rm skin}$ \\
\hline \hline
NL3$^{\ast}$      & 24.9 $\pm$ 0.4 & 30.7 $\pm$ 0.7 & 50.3 $\pm$ 1.8 &\phantom{xx}39.2 $\pm$ 17.8  & -284.6 $\pm$ 29.4  & 0.18 $\pm$ 0.01\\
SkNL3$^{\ast}$    & 24.5 $\pm$ 0.3 & 31.0 $\pm$ 0.9 & 46.4 $\pm$ 6.4 & \phantom{x}62.7 $\pm$ 16.6   & -380.0 $\pm$ 15.2  & 0.16 $\pm$ 0.01\\
IU-FSU            & 24.9 $\pm$ 0.4 & 31.4 $\pm$ 0.7 & 52.9 $\pm$ 2.0 &\phantom{xx}-6.8 $\pm$ 12.9   & -257.6 $\pm$ 22.3  & 0.18 $\pm$ 0.01 \\
SkIU-FSU          & 24.4 $\pm$ 0.3 & 31.4 $\pm$ 0.9 & 48.0 $\pm$ 6.2 &-130.2 $\pm$ 13.3  & -343.9  $\pm$ 15.3 & 0.16  $\pm$ 0.01 \\
\hline
\end{tabular}
\end{center}
\end{table}

%%%%%%%%%%%%%%%%%%%%%%%%%%%%%%%%%%%%%%%%%%%%%%%%%%%%%%%%%%%%%%%%%
%%%%%%%%%%%%%%%%%%%%%%%%%%%%%%%%%%%%%%%%%%%%%%%%%%%%%%%%%%%%%%%%%

Particularly attractive are predictions for the neutron skin
thickness of $^{208}Pb$ after the PNM optimization. The original
NL3* parametrization predicts a very thick neutron skin of $R_{\rm
skin} = 0.29$ fm, while its Skyrme counterpart predicts a value of
$R_{\rm skin} = 0.27$ fm. On the other hand, both the original
IU-FSU and SkIU-FSU parameterizations predict a much lower value of
$R_{\rm skin} = 0.16$ fm for $^{208}$Pb. These values are consistent
with the current experimental result of $R_{\rm skin} =
0.33^{+0.16}_{-0.18}$ fm for the neutron skin thickness of lead
obtained using electroweak probes in the PREX
experiment~\cite{Abrahamyan:2012gp}. After the PNM optimization, we
find that in general the RMF model predicts $R_{\rm skin} = 0.18 \pm
0.01$ fm for $^{208}$Pb, while the SHF model predicts $R_{\rm skin}
= 0.16 \pm 0.01$ fm (See Table \ref{Table3}), which are marginally
consistent with each other within the 1$\sigma$ error-bars. Thus
both models generically predict a thin neutron skin thickness of
lead. Moreover, the error-bars coming from this constraint are very
small. This is a particularly provocative result, since if the new
PREX experiment reduces the error bars without moving the central
value for the neutron skin, almost all current models of the nuclear
structure would need to be modified. Also, this would appear to call
for a significant modification of the PNM microscopic calculations.

%%%%%%%%%%%%%%%%%%%%%%%%%%%%%%%%%%%%%%%%%%%%%%%%%%%%%%%%%%%%%%%%%
\begin{figure}[h]
\vspace{-0.05in}
\begin{center}
\includegraphics[width=0.85\columnwidth,angle=0]{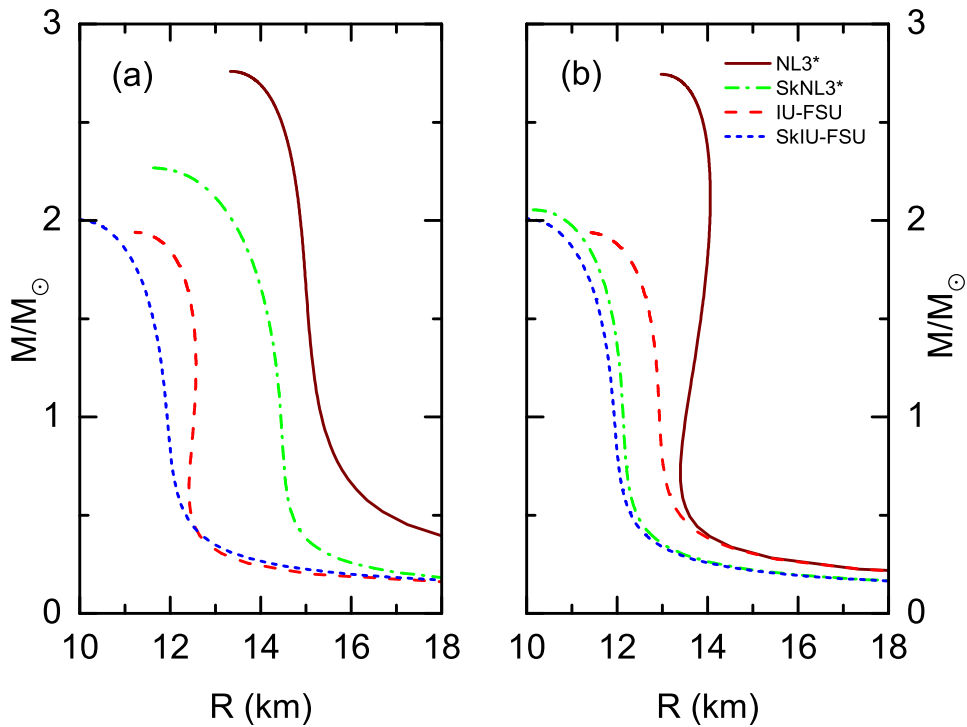}
\caption{Mass-vs-radius relation of neutron stars calculated from
the four parameterizations before (a) and after (b) the PNM
optimization (Figure is taken from Ref.~\cite{Fattoyev:2012b}).}
\label{Fig3}
\end{center}
\end{figure}
%%%%%%%%%%%%%%%%%%%%%%%%%%%%%%%%%%%%%%%%%%%%%%%%%%%%%%%%%%%%%%%%%

%%%%%%%%%%%%%%%%%%%%%%%%%%%%%%%%%%%%%%%%%%%%%%%%%%%%%%%%%%%%%%%%%
\begin{figure}[h]
\vspace{-0.05in}
\begin{center}
\includegraphics[width=0.75\columnwidth,angle=0]{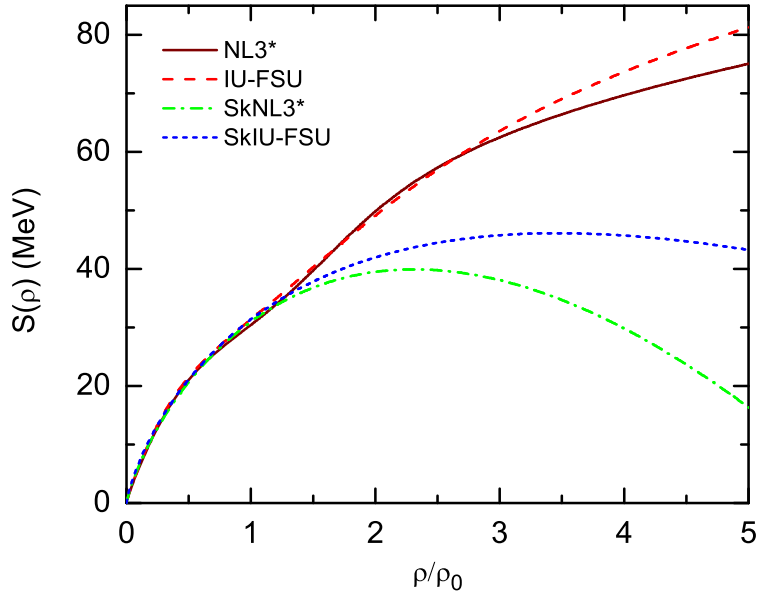}
\caption{Density dependence of symmetry energy from the four
parameterizations after the PNM optimization (Figure is taken from
Ref.~\cite{Fattoyev:2012b}).} \label{Fig4}
\end{center}
\end{figure}
%%%%%%%%%%%%%%%%%%%%%%%%%%%%%%%%%%%%%%%%%%%%%%%%%%%%%%%%%%%%%%%%%

Finally, we apply our results to the neutron star structure, by
extrapolating the EoS from the RMF and the SHF models to higher
densities. The neutron star matter is assumed to be charge neutral
and in the $\beta$-equilibrium condition with neutrons, protons,
electrons, and muons. No exotic degrees of freedom are assumed. The
equations of state from the four parametrization are then utilized
in the general relativistic equation of stellar structure (known as
the Tolman-Oppenheimer-Volkoff equation) to obtain the mass-{\sl
vs}-radius relation. In Fig. \ref{Fig3} (a) we display mass-{\sl
vs}-radius relations as predicted by the four original RMF and SHF
parametrizations that predict a wide range of results for low mass
neutron star radii. This can be mainly attributed to the density
dependence of the symmetry energy, which is quite different in the
original two parameterizations. Once calibrated to the PNM results,
this difference almost vanishes within the same model as shown in
the Fig. \ref{Fig3} (b), i.e., both RMF and SHF parameterizations
now match each other more closely. The only difference at high
masses between the RMF parametrizations is now due to the $\zeta$
parameter that results in a stiffer EoS of SNM in NL3$^{\ast}$
parameterizations at several times saturation density. Although both
NL3$^{\ast}$ and IU-FSU parameterizations in a given RMF or SHF
model predict similar radii, there is a clear difference between the
RMF and the SHF predictions as a whole. In the case of IU-FSU and
SkIU-FSU we have almost a $\sim 1$ km difference for the radius of a
canonical neutron star. This discrepancy is even larger in the case
of NL3$^{\ast}$, which is about $\sim 1.8$ km. Thus, there is a
strong model dependence when the two models are applied to neutron
star structure calculations after the same PNM optimization. This
model dependence primarily comes from the different density
dependence of symmetry energy at supra-saturation densities. Looking
at Fig. \ref{Fig4} we see that the symmetry energy is almost the
same in all the models up to $\sim 1.5\rho_0$. The RMF model
predicts a monotonic increasing function of density for the symmetry
energy, while the SHF model tends to give a decreasing symmetry
energy at higher densities. We thus show that the low-density PNM
constraints alone result in a distinct model dependency of radius
predictions. With the similar saturation properties of nuclear
matter constrained by the PNM EoS, one can obtain different radii
for a given neutron star mass. Although the PNM optimization tightly
constrains the symmetry energy up to a little above the saturation
density, its high density behavior that is very crucial in
determining neutron star radii, still remain unclear. To elucidate
this long-standing problem further we need to rely on the heavy-ion
collision
experiments~\cite{Danielewicz:2002pu,Li:2008gp,Baran:2004ih} and
neutron star observations~\cite{Lattimer:2006xb, Steiner:2004fi}.

\section{Summary}
\label{summary}

Using our best knowledge of the PNM EoS below and around saturation
density from {\it ab initio} calculations, we constrain the density
dependence of the symmetry energy for the RMF and SHF models, by
optimizing the two pure isovector parameters from each model while
keeping the values of other parameters so that the errors of
predicted binding energies and charge radii of medium to heavy
nuclei remain to be less than $2\%$.

We show that such fits result in very similar predictions for the
symmetry energy $J$ and its slope parameter $L$ at saturation
density from both models as long as the nucleon effective mass from
both RMF and SHF models is chosen to be
consistent~\cite{Jaminon:1989wj, Furnstahl:1997tk}. When the error
bounds are plotted as ellipses in the $J$-$L$ plane, a
positively-correlated relationship between $J$ and $L$ is observed
for both models. However, different slopes are obtained from the RMF
and SHF models, and the two ellipses have no overlapping area in the
plane. This model dependence comes from the different values of
$K_{\rm sym}$ and higher-order symmetry energy parameters.

Predictions of a neutron skin thickness $R_{\rm skin}$ for
$^{208}$Pb are similar from both models and are within the error-bar
of the latest experimental data. Both models predict a very thin
neutron skin. Although the PNM constraints lead to broadly similar
behaviors of the symmetry energy as a function of density up to
$\approx 1.5 \rho_0$, they deviate significantly at higher densities
due to the differences in the functional form of the symmetry
energy. This results in a striking difference in the predictions of
the neutron star radii from both models.

We showed the possible differences in predictions from two
phenomenological energy-density functional forms, when the same
experimental or theoretical constraints up to saturation density are
applied. Care must be taken interpreting observational and
experimental constraints from different nuclear models, and
searching for a robust and better-determined EDF is necessary.

\section{Acknowledgments}
This work is supported in part by the National Aeronautics and Space
Administration under grant NNX11AC41G issued through the Science
Mission Directorate, and the National Science Foundation under
Grants No. PHY-1068022 and No. PHY-0757839.

\end{document}